# Tuning the Interlayer Microstructure and Residual Stress of Buffer-Free Direct Bonding GaN/Si Heterostructures

Yan Zhou[1,2,a),b)], Shi Zhou[3,a)], Shun Wan[4,a),b)], Bo Zou[5], Yuxia Feng[6], Rui Mei[2], Heng Wu[2], Pingheng Tan[2], Naoteru Shigekawa[7,8], Jianbo Liang[7,8,b)] and Martin Kuball[1,b)]

[1] *Center for Device Thermography and Reliability (CDTR), H. H. Wills Physics Laboratory, University of Bristol, Tyndall Avenue, Bristol BS8 1TL, UK.*

[2] *State Key Laboratory of Superlattices and Microstructures, Institute of Semiconductors, Chinese Academy of Sciences, Beijing 100083, China.*

[3] *Nano Science and Technology Institute, University of Science and Technology of China, Hefei 230026, China.*

[4] *Center for High Pressure Science and Technology of Advanced Research, Shanghai, 201203, China.*

[5] *School of Science, Harbin Institute of Technology, Shenzhen 518055, China.*

[6] *Key Laboratory of Optoelectronics Technology, Ministry of Education, Beijing University of Technology, Beijing, 100124, China*

[7]*Department of Electronic Information Systems, Osaka City University, Sugimoto 3-3-138 Sumiyoshi, Osaka 558-8585, Japan*

[8]*Graduate School of Engineering, Osaka Metropolitan University, Sugimoto 3-3-138, Sumiyoshi, Osaka 558-8585, Japan*

[a)]Co-first authors, contributed equally. [b)]Corresponding authors. Email address: yan.zhou@bristol.ac.uk, shun.wan@hpstar.ac.cn, liang@omu.ac.jp, martin.kuball@bristol.ac.uk.

**Abstract:** The direct integration of GaN with Si can boost great potential for low-cost, large-scale, and high-power device applications. However, it is still challengeable to directly grow GaN on Si without using thick strain relief buffer layers due to their large lattice and thermal-expansion-coefficient mismatches. In this work, a GaN/Si heterointerface without any buffer layer is successfully fabricated at room temperature via surface activated bonding (SAB). The residual stress states and interfacial microstructures of GaN/Si heterostructures were systematically investigated through micro-Raman spectroscopy and transmission electron microscopy. Compared to the large compressive stress that existed in GaN layers grown-on-Si by MOCVD, a significantly relaxed and uniform small tensile stress was observed in GaN layers bonded-to-Si by SAB; this is mainly ascribed to the amorphous layer formed at the bonding interface. In addition, the interfacial microstructure and stress states of bonded GaN/Si heterointerfaces was found can be significantly tuned by appropriate thermal annealing. With increasing annealing temperature, the amorphous interlayer formed at the as-bonded interface gradually transforms into a thin crystalline silicon nitride interlayer without any observable defects even after 1000˚C annealing, while the interlayer stresses at both GaN layer and Si monotonically change due to the interfacial re-crystallization. This work moves an important step forward directly integrating GaN to the present Si CMOS technology with high quality thin interfaces, and brings great promises for wafer-scale low-cost fabrication of GaN electronics.

Gallium nitride (GaN) is one of the most important semiconductor materials for present and future high-power and high-frequency devices,[1-3] mainly due to its superior physical properties of wide bandgap, high electron saturation velocity, high breakdown field, and high chemical stability.[4-6] Though there are many commercial prospects to use SiC or diamond as substrates for GaN, one of the critical issues GaN-based devices still face is their comparably expensive cost for compatible with Si-based integrated circuits. Integrating GaN with Si substrates would enable the fabrication of low-cost and multi-functional chips.[7] Although GaN epitaxial layers integrated on Si through crystal growth methods has become a commercial product, thick strain relief layers such as AlN, AlN/AlGaN, AlN/GaN need to be used because of the large residual stresses induced by their large mismatches in lattice (17% between GaN and Si) and thermal-expansion-coefficient.[8-15] Large residual stresses affect the electrical and optical properties of GaN-based devices.[16, 17] In addition, the strategy of growing thick strain relief layers for GaN/Si heterostructures also introduces a large thermal boundary resistance at interfaces, thus degrading the critical thermal management performance of GaN-based devices.[10, 18, 19]

To realize high quality, low stress GaN/Si heterointerfaces, integrating GaN with Si via wafer bonding has attracted tremendous interests and has achieved great progress through using various buffer layers;[20-22] however, the direct bonding of GaN to Si substrates without any buffer layers is still rather challenging though very attractive as it would remove any sizable thermal barrier at the GaN/Si interface. Recently, room-temperature direct bonding of dissimilar materials with large lattice mismatch between heterostructures such as diamond/Si, GaN/diamond, GaN/SiC via surface-activated-bonding (SAB) exhibited excellent performances and great potential.[23-26] Despite a relative thick amorphous interlayer is normally formed at the interface after SAB,[23-26] which can affect the residual stress, interlayer microstructure and device performance of various heterostructures that is still not fully understood, thermal annealing is known can modulate the defects, modify the amorphous structure and change the residual stress of heterostructures.[23, 24, 27, 28] Though in-situ measuring the interlayer microstructure and residual stress during annealing is still a challenge, confocal micro-Raman spectroscopy can provide sub-micrometer resolution to reveal the local stress.[23, 24, 29-34] The atomic-scale microstructure of various heterointerfaces after different annealing can be revealed through transmission electron microscopy (TEM).[23, 24]

In this work, direct bonding of GaN/Si heterostructures without any buffer layers were realized here by SAB at room temperature. The residual stress and the interlayer microstructure along with the annealing impacts of GaN/Si heterostructures were systematically investigated and compared with that of MOCVD grown GaN/Si heterostructures, using confocal micro-Raman spectroscopy and TEM. This work will provide a new perspective for controlling the residual stress and interface quality in GaN/Si or similar heterostructures, and for developing novel device structures.

GaN(0001) epitaxial layers used in this study were grown by MOCVD on 4-inch n-Si(111) substrates. To grow the GaN/Si heterostructures, trimethylgallium (TMG), trimethyl-aluminum (TMA), and ammonia ($NH_3$) were used as sources for Ga, Al, and N, respectively. Prior to grow GaN, the Si substrate was cleaned in $H_2$ ambient at 1000 °C, followed by the growth of about 200 nm-thick AlN buffer layer at 1060 °C. A 1.2 μm-thick GaN layer was then grown on the AlN buffer layer at 1000 °C. The GaN epitaxial layers grown-on-Si and an additional target n-Si(111) substrate used for the bonding were cleaned with acetone and ethanol in an ultrasonic bath for 300 s, dried under $N_2$, and then loaded into the SAB chamber. After their surfaces were activated by fast Ar atom beam irradiation,[35, 36] the GaN epitaxial layers were bonded to the target n-Si(111) substrates at room temperature under optimized bonding parameters of a vacuum pressure of 5 x $10^{-7}$ Pa and an external load of 1 GPa for 60 s, resulting in a Si/GaN/Si heterostructure. After bonding, the grown Si substrate was removed by mechanical polishing and chemical wet-etching. To avoid etching the bonded n-Si(111) substrate, a $SiO_2$ layer was deposited on the backside of the Si substrate by the RF sputtering. Here the grown GaN-on-Si is Ga-polar, while the SAB process results in an N-polar GaN-on-Si. The detailed process of GaN/Si heterostructure fabricated by SAB is shown in Fig. 1.

The residual stress in the SAB fabricated GaN/Si heterostructure was investigated systematically by micro-Raman spectroscopy. Raman mapping measurements were confocal on the GaN surface and the interface to the bonded n-Si(111) wafer in an area of 40×40 μm$^2$ mapped with a step size of 1 μm. A Renishaw InVia confocal micro-Raman spectroscopy with a measurement configuration of 180° backscattering geometry, an $Ar^+$ laser of 488 nm, a diffraction grating of 2400 lines/mm, a 50×0.6NA long working distance objective lens, and a laser sport size of ~1.5 μm was used. Samples were also measured at elevated temperatures by in situ micro-Raman in a high-temperature heating/cooling stage (Linkam TS-1500) with a temperature stability of

±1 °C for controlling the sample temperature. The bonded samples were annealed at 400, 700, and 1000 °C for 300 s in $N_2$ gas atmosphere. The GaN and Si Raman mode frequencies were determined by performing a Lorentz fitting to their recorded spectra, with a shift resolution as low as ±0.02 cm$^{-1}$ within the measurement time; Raman shift of the Si calibration wafer before and after each measurement was measured to check for stability and accuracy. The $E_2^{high}$ mode of GaN was employed to monitor the residual stress in GaN layers because of its higher stress sensitivity.[37] The interfacial microstructures of GaN/Si heterostructure were investigated by TEM (JEM-2200FS).

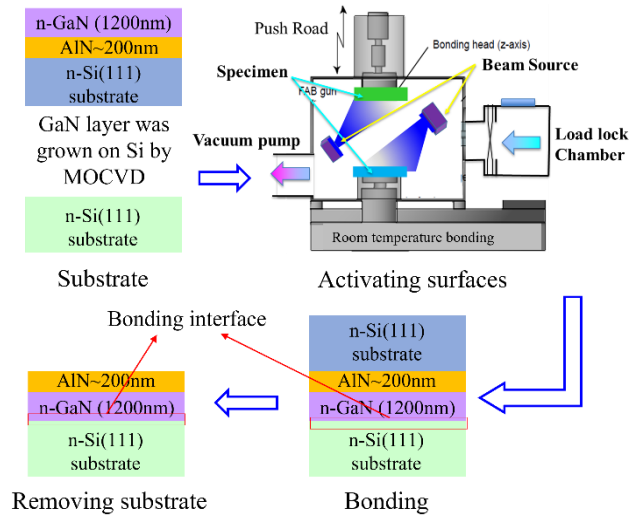

**FIG. 1.** Schematic process of fabricating a GaN/Si heterostructure using SAB.

Raman peak positions from mapping the GaN $E_2^{high}$ mode in SAB fabricated GaN/Si heterostructures and in the MOCVD as-grown sample prior to SAB are shown in Figs. 2(a-b); corresponding histograms of peak positions and stress distributions are displayed in Figs. 2(c-d). For unstressed GaN, numerous values for Raman peak of GaN $E_2^{high}$ mode have been reported in the literature.[38-40]; here, we measured a 356 μm-thick free-standing GaN substrate grown by hydride vapor-phase-epitaxy as a reference stress-free bulk value, determining the Raman peak of GaN $E_2^{high}$ at 567 cm$^{-1}$. For the Si $F_{2g}$ mode in stress-free case, 520 cm$^{-1}$ was measured. As shown in Figs. 2(a-b), the Raman peak of the GaN $E_2^{high}$ mode in GaN/Si heterostructures fabricated by SAB and MOCVD was found to range from 566.57 to 567.46 cm$^{-1}$ and 566.35 to 567.87 cm$^{-1}$; their averaged Raman peak positions were determined to be 567±0.13 and 567.14±0.25 cm$^{-1}$, respectively (Figs. 2(c-d)). The GaN peak shifted to higher wave numbers compared to the stress-free bulk value of 567 cm$^{-1}$ indicates that the GaN layer grown-on-Si was under compressive stress. Note that there were some variations in the GaN peak of MOCVD grown GaN/Si heterostructure, indicating some inhomogeneity existed. According to literatures, the phonon frequencies of GaN $E_2^{high}$ mode and Si $F_{2g}$ mode both shift linearly with stress at a rate of 2.9 and 2.3 cm$^{-1}$ GPa$^{-1}$, respectively.[41, 42] We used these coefficients to calculate the residual stress values in GaN/Si heterostructures from the Raman peak shifts of GaN and Si with respect to their stress-free bulk values. As demonstrated in Figs. 2(c-d), a near-zero small tensile residual stress of 0±0.05 GPa with respect to the unstressed bulk value was extracted from the GaN layer of the SAB fabricated GaN/Si heterostructures, implying the residual stress is significantly relaxed in SAB as-bonded GaN/Si heterostructures, while a compressive stress of -0.05±0.09 GPa was present in the GaN layer of MOCVD grown GaN/Si heterostructures; this is consistent with values reported in the literature for GaN grown on Si (111) by MOCVD using similar strain relief buffer layers.[39, 43, 44]

Residual stress in SAB fabricated heterostructures were previously found to change with annealing temperature.[23, 24] Based on these earlier studies, we explored the effects of thermal annealing on residual stress and interfacial structure of SAB fabricated GaN/Si heterostructures. The averaged residual stresses and corresponding statistic distributions in GaN and Si layers of SAB fabricated GaN/Si heterostructures after annealing at different temperatures are shown in Figs. 2(e-f), respectively; while those in GaN and Si layers of MOCVD grown GaN/Si heterostructures were also measured at room temperature for comparison in Figs. 2(e-f). Here, the shown residual stresses represent the averaged value over the area of 40 × 40 μm$^2$. The magnitude of error bars and the nearby statistic curves reflected the range and homogeneity of stress distribution in GaN epilayer and Si substrate. Residual stresses along with their distribution ranges of GaN and Si in SAB fabricated GaN/Si heterostructures prior to annealing were smaller than those of MOCVD grown GaN/Si heterostructures. Moreover, the average residual stresses of GaN and Si change to different levels with increasing annealing temperature at different rates, for example, after annealing at 1000 °C, their residual stresses become a larger tensile (0.16 GPa) and a smaller compressive (0.06 Gpa) stress, respectively. However, below 700 °C annealing, there is negligible changes in the residual stress of GaN, while the compressive stress in Si decreased linearly with increased temperature.

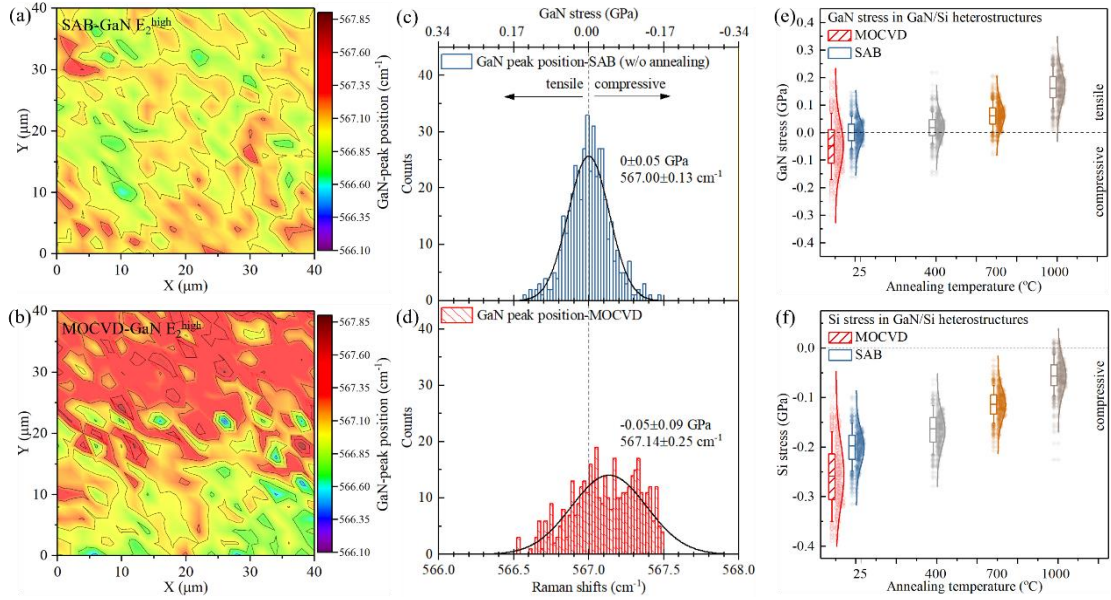

**FIG. 2.** Raman peak position maps of $E_2^{high}$ mode of GaN layer in (a) GaN/Si heterostructures fabricated by SAB and (b) MOCVD as-grown sample on Si substrate. Histogram statistic distributions for Raman peak positions and the residual stresses of GaN in corresponding structures extracted from their Raman data are shown in (c) and (d), respectively; a Raman peak of 567 cm$^{-1}$ is taken for unstressed GaN as the reference value while 2.9 cm$^{-1}$/GPa is taken for its stress coefficient. Red and blue colors represent MOCVD and SAB samples, respectively. Residual stresses of GaN (e) and Si (f) in SAB fabricated GaN/Si heterostructures, averaged over an area of 40×40 μm$^2$, after annealing at different temperatures. Error bars represent the standard deviation of the average value and reflect the level of homogeneousness of residual stress.

To gain insight into the interfacial microstructure evolution under different annealing temperatures, the local nanoscale and atomic-scale SAB bonded GaN/Si interfaces were examined by TEM. As shown in Figs. 3(a-d), low and high magnification, as well as high-resolution cross-sectional TEM images of the GaN/Si bonding interface fabricated by SAB without and with annealing at 1000 °C were performed respectively. It is apparent in Fig. 3(a) that, a relatively sharp interface without any micro voids were observed at the bonding interface, including after removal of the growth substrate. Meanwhile, a disordered amorphous-like interlayer with a thickness of tens of nm was present at the un-annealed bonding interface (Fig. 3(b)). After annealing at 1000 °C, no amorphous layer was observed at the bonding interface (Fig. 3(c)); instead, a crystallized interlayer of only few nm thin was formed at the bonding interface (Fig. 3(d)). More importantly, no structural defects such as cracks were observed at the interface whether without or with annealing. Further EDS mappings of the un-annealed and 1000 °C annealed bonding interface show that the crystallized interlayer is composed of Ga, N and Si elements (Fig. 3(e-f)). We also note that after 1000 °C annealing, the intensity gradients of Ga, N, and Si show Si atoms diffused into the GaN side, and Ga and N atoms diffused into the Si adjacent to the bonding interface due to the thermal annealing effects.

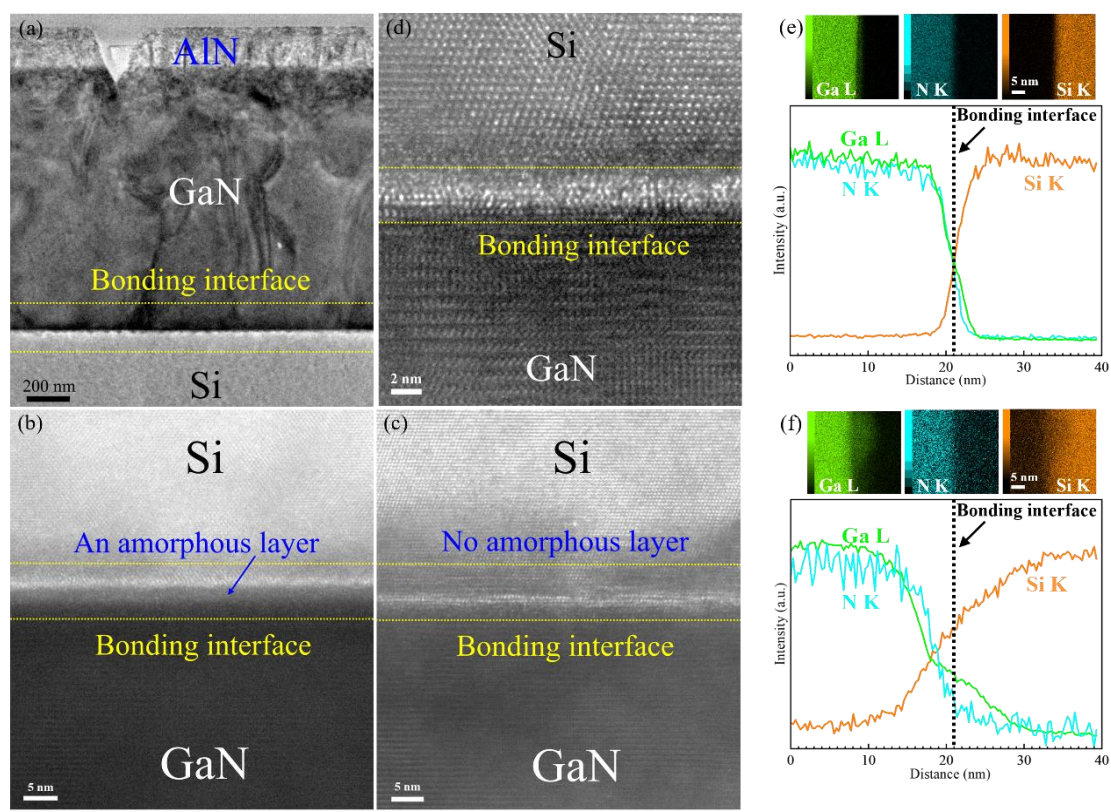

**FIG. 3**. A low magnification (a) and high magnification (b) cross-sectional TEM images of the GaN/Si bonding interface fabricated by SAB without annealing, as well as a high magnification (c) and high resolution (d) cross-sectional TEM images of the GaN/Si bonding interface fabricated by SAB after annealing at 1000 °C. The EDS mapping of the SAB bonded interface region (e) without annealing and (f) after annealing at 1000 °C.

Relaxation of residual stress occurred in the GaN layer of SAB fabricated GaN/Si heterostructure should be attributed to the formation of an amorphous layer at the bonding interface. Although a similar amorphous layer was also observed at the SAB fabricated Si/SiC, Si/GaAs, Si/Diamond interfaces,[23, 25, 45, 46] this amorphous layer plays a cushioning effect across the GaN/Si bonding interface, which can relax the residual stress caused by the large difference in the lattice constant between Si and GaN (~17%). The formation of the amorphous layer at the bonding interface is due to the fast Ar atom beam activation during the bonding process. It was found that the residual stress in SAB bonded GaN/Si heterostructure strongly depended on the post-annealing temperature. Also, no sizable amorphous layer was observed at the bonding interface after 1000 °C annealing; moreover, the TEM images (Figs. 3) illustrate that the amorphous layer recrystallized after annealing at high temperature. The residual stress in GaN of SAB fabricated GaN/Si heterostructures increased with increasing annealing temperature, which should be correlated with the decrease in the amorphous layer thickness, making it more and more difficult to ease up the residual stress caused by lattice and thermal-expansion mismatches between GaN and Si. Because the lattice-constant of GaN is smaller than that of Si (111) and its thermal-expansion-coefficient is larger than that of Si, therefore, tensile stress becomes more significant in the GaN layers with increasing annealing temperature. While the compressive stress in Si near the bonding interface decreased with increasing annealing temperature.

It has been reported that a silicon nitride layer can form at the GaN/Si interface due to the direct reaction of nitride with silicon.[47] We are then led to the hypothesis that a compressive stress could exist in the interlayer if the silicon nitride layer was to be formed at the GaN/Si bonding interface. Because there is a large residual stress difference between GaN and Si layers near the bonding interface after 1000 °C annealing, consequently, a compressive stress likely exists in the recrystallized silicon nitride interlayer to compensate the reduction of compressive stress in Si. Moreover, we note no structural defects were observed at the bonding interface. These advantages indicated that it is possible to obtain stress-free GaN epitaxial layers through SAB technique at room temperature, and tune the interlayer structure and residual stress through appropriate temperature annealing.

In conclusion, GaN epitaxial wafers were directly bonded to Si(111) substrates by SAB without any buffer at room temperature. The residual stress and interlayer microstructure in SAB bonded GaN/Si heterostructure were systematically investigated and compared with that of MOCVD grown GaN/Si heterostructure, using confocal micro-Raman spectroscopy and TEM. It was found that the residual stress was greatly relaxed in SAB fabricated GaN/Si heterostructures and a uniform distribution of small tensile residual stress was obtained; these contrasts to a larger compressive stress in MOCVD grown GaN/Si heterostructures. The main reason was an amorphous layer formed at the bonding interface supporting stress relaxation. The interlayer microstructure and residual stress of SAB bonded GaN/Si heterostructures could be significantly tuned by appropriate thermal annealing. Residual stress was observed monotonically change with increasing temperature, correlating with the evolution of amorphous layer. After 1000˚C annealing, the amorphous interlayer disappeared while a few nm thin silicon nitrite interlayer formed at the bonding interface, without any observable defects. This work realizes the buffer-free direct bonding of GaN epilayer with Si substrate, unveils the tuning potential of GaN/Si interfaces, and provides great prospects for low-cost, large-scale, and multi-functional GaN/Si device applications.

**Acknowledgements:**

This work was in part supported by JSPS KAKENHI under grant number JP20K04581, the Osaka City University (OCU) Strategic Research Grant 2020 for top basic research, and the National Natural Science Foundation of China under grant number 42050203, and China Postdoctoral Science Foundation under grant number 2019M663904XB. Any opinions, findings, and conclusions or recommendations expressed in this material are those of the authors and do not necessarily reflect the views of JSPS.

**Conflict of Interest:**
The authors declare no conflict of interest.

**Data Availability Statement:**
The data that supports the findings of this study are available within the article, or from the corresponding author upon reasonable request.